\documentclass[%
 reprint,
aps,
prl,
floatfix,
]{revtex4-2}

\usepackage{graphicx}% Include figure files
\usepackage{dcolumn}% Align table columns on decimal point
\usepackage{bm}% bold math
\begin{document}

\preprint{APS/123-QED}

\title{Analytical model and experimental validation for \\nonlinear mechanical response of aspirated elastic shells  
}% Force line breaks with \\

\author{Kazutoshi Masuda}
 \email{kazumassuu@g.ecc.u-tokyo.ac.jp}
\author{Miho Yanagisawa}%
 \email{myanagisawa@g.ecc.u-tokyo.ac.jp}
 \altaffiliation[Also at ]{Komaba Institute for Science, Graduate School of Arts and Sciences, The University of Tokyo, Komaba 3-8-1, Meguro, Tokyo 153-8902, Japan; Center for Complex Systems Biology, Universal Biology Institute, The University of Tokyo, Komaba 3-8-1, Meguro, Tokyo 153-8902, Japan; Department of Physics, Graduate School of Science, The University of Tokyo, Hongo 7-3-1, Bunkyo, Tokyo 113-0033, Japan; }%Lines break automatically or can be forced with \\
\affiliation{%
 Department of Basic Science, Graduate School of Arts and Sciences, The University of Tokyo, Komaba 3-8-1, Meguro, Tokyo 153-8902, Japan
}%

\date{\today}% It is always \today, today,

\begin{abstract}
We developed a physics-based analytical model to describe the nonlinear mechanical response of aspirated elastic shells. By representing the elastic energy through a stretching modulus, \(K\), and a dimensionless
ratio, \( \delta \), capturing the balance between stretching and bending energies, the model reveals mechanical behaviors extending beyond conventional approaches. Validated across microscale droplets and macroscale silicone sheets by fitting experimental force-displacement curves, this approach provides accurate, scalable characterization of deformed elastic shells. This framework advances our understanding of soft thin-shell mechanics, with broad applications in probing living cells and designing soft materials.
%\begin{description}
%\end{description}
\end{abstract}

\maketitle

%\tableofcontents
The mechanics of soft materials, such as polymer droplets, lipid vesicles, and living cells, play a crucial role in both fundamental science and practical applications \cite{DeGennes1992, Doi2013, Levental2007}. To characterize their mechanical properties, various experimental techniques have been utilized \cite{Bao2003, Wubshet2023}, including atomic force microscopy \cite{Thomas2013, Kis2002}, optical tweezers \cite{Furst1999, Ou-Yang2010}, and micropipette aspiration \cite{Hochmuth2000, Tian2007}. These conventional methods enable quantitative measurements of force-displacement (or stress-strain) relationships and estimate mechanical parameters such as Young’s modulus \cite{Theret1988, Lytra2020}, shear modulus \cite{Evans1973, Waugh1979, Hnon1999} and viscosity \cite{Guevorkian2010} in both static and dynamic regimes. However, these analyses often assume material homogeneity and linear elasticity, which limits their relevance to complex soft matter systems.

In particular, capsule-like systems, such as polymer-coated gel capsules and living cells, exhibit mechanical heterogeneity and nonlinear force responses significantly even in the absence of shear due to their thin-shell structures \cite{Waugh1979, Esteban-Manzanares2017}. These nonlinearities are poorly captured by conventional methods, yet the linear approximations are still widely applied, even if limited to a small initial range \cite{Sakai2018}. As a result, Young’s modulus derived from such models often lack reliability and fail to explain key behaviors, such as deformation through narrow lumens or in confined geometries. Although various empirical models and simulation works have been proposed to account for these behaviors, they typically lack a clear connection to the underlying physics or do not reproduce the full nonlinear response observed in experiments \cite{Rudenko2014, Esteban-Manzanares2017, Ourique2022, Du2023}.

In this study, we present a physics-based analytical model that quantitatively describes the nonlinear mechanical response of aspirated elastic shells. This model couples bending and in-plane stretching deformations, which have not been simultaneously captured in previous studies as shown in \cite{Landau2012, Rawicz2000}. 
We extract two physically meaningful parameters, (i) the area stretching modulus, \(K\), and (ii) the dimensionless bending contribution ratio, \(\delta\), by fitting to the experimentally measured force-displacement curves.

\begin{figure}
\includegraphics[width=1 \linewidth] {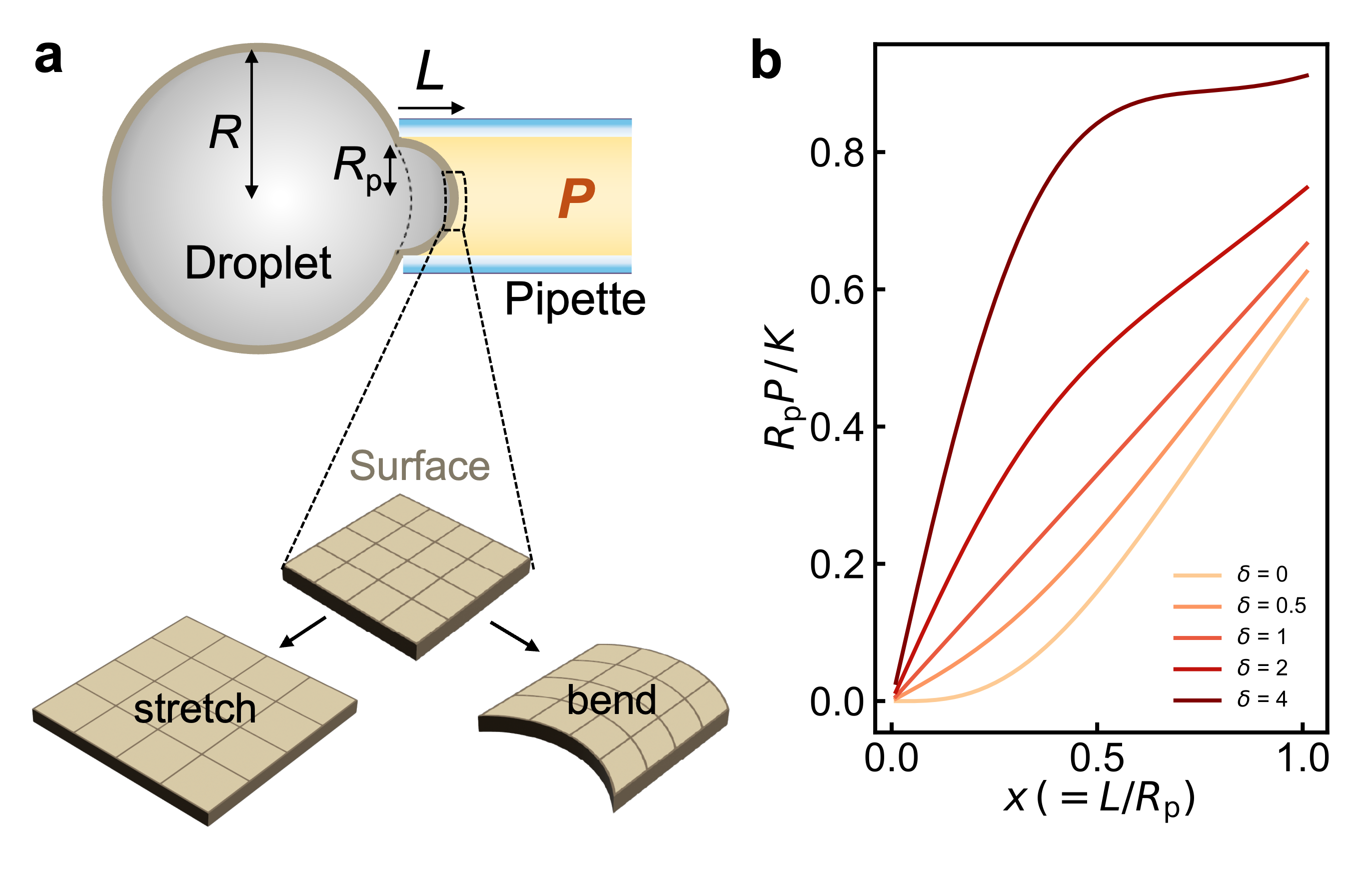}
\caption{\label{fig:model} Analytical model for the deformation of membrane-covered spherical droplets during micropipette aspiration. (a) A droplet with radius \(R\) is aspirated by a pipette with an inner radius \(R_{\mathrm{p}}\) under a certain pressure difference \(P\) between the external environment and the interior of the pipette. The resulting deformation in the aspirated region includes stretching the surface area due to the aspiration length $L$ and bending of the surface. (b) Plot showing the numerically obtained relationship between the normalized pressure $R_{\mathrm{p}}P/K$ and the normalized aspiration length, $x$ $(=L/R_{\mathrm{p}})$ for various values of \(\delta\). The curve becomes linear when \(\delta=1\); it becomes convex upwards when \(\delta>1\) and and convex downwards \(\delta<1\).}
\end{figure}

We consider the classical setup of micropipette aspiration applied to a spherical droplet coated with an elastic membrane, as illustrated in Fig.~\ref{fig:model}(a). In this configuration, a pressure difference \(P\) between the interior and exterior of the pipette induces deformation of the membrane, which is aspirated into the pipette by a length \(L\). We focused on the regime of small deformations, where the normalized aspiration length \(x = L/R_{\mathrm{p}}\) (with \(R_{\mathrm{p}}\) the pipette radius) remains below unity. In this regime ($x < 1$), deformations outside the pipette can be neglected, and the mechanical response is governed primarily by the geometry and energies of the aspirated regions.

We then analyzed the changes in the elastic energies consisting of the bending energy by interfacial curvature difference, $E_b$, and the stretching energy due to the surface area expansion, $E_a$ to obtain $P\text{-}x$ fitting;
\begin{equation}\label{eq:total}
E_a + E_b = \int P(x) \frac{dV}{dx} \, dx.
\end{equation}

\noindent
The volume \(V\) and surface area of the aspirated region \(A\) can be expressed geometrically as follows:
\begin{equation}\label{eq:v}
V = \frac{1}{6} \pi R_{\mathrm{p}}^3 (3x + x^3) \>,
\end{equation}
\begin{equation}\label{eq:A}
A =\pi R_{\mathrm{p}}^2 (1 + x^2).
\end{equation}

The bending energy, $E_b$ is expressed with a bending rigidity, $k$ and the curvature radius of the aspirated region, \( R^{\prime} \) as follows:
\begin{equation}\label{eq:bE}
E_b = \frac{1}{2} k \left( \frac{2}{R^{\prime}} \right)^2 A = 8 \pi k \frac{x^2}{1 + x^2},
\end{equation}
where \( R^{\prime} \) and \textit{A} are given by \(R' = R_{\mathrm{p}} (1 + x^2)/2x \) and Eq.~\ref{eq:A}, respectively. To simplify the calculation, the bending energy before aspiration was set to zero.

The stretching energy, $E_a$ is expressed from the interfacial energy of a membrane composed of molecules such as lipids. The interfacial energy per molecule \( E_a^{\text{single}}\) is given as follows.
\begin{equation}\label{eq:2}
E_a^{\text{single}} = \frac{\alpha}{a} + K \left( \frac{a}{a_0} \right)^2 a_0 ,
\end{equation}
where $a$ and $a_0$ are the apparent area occupied by a single molecule and its equilibrium value, respectively. \textit{K} is the area stretching modulus, which is known to play a vital role in many biological \cite{DEDischer1994, Suresh2005, Han2020, Linke2024, Massey2024} and polymeric \cite{Yu2024} systems. 
\( \alpha \) is a repulsive force constant due to molecular rearrangement as the area increases \cite{Israelachvili2011}, which is determined from the equilibrium condition ($dE_a^{\text{single}}/da=0$) as \( \alpha = 2 K a_0^2 \).

Considering the short aspiration time (within a few minutes) to ignore the influx of new molecules from outside, the number of molecules in the aspirated area is constant as follows;
\begin{equation}\label{eq:3}
\frac{A_0}{a_0} = \frac{A}{a},
\end{equation}
where \( A_0 \) represents the initial aspirated area.

By multiplying the number of molecules present in the aspirated region by $E_a^{\text{single}}$ using Eqs.~\ref{eq:2},\ref{eq:3}, the stretching energy $E_a$ in the aspirated region is derived as follows;

\begin{equation}
E_a = \frac{A_0}{a_0} \cdot E_a^{\text{single}} = K \left[ \frac{A^2}{A_0} + \frac{2A_0^2}{A} \right],
\end{equation}

\noindent and further refined by using Eq.~\ref{eq:A} as:
\begin{equation}\label{eq:aE}
E_a = \pi K R_{\mathrm{p}}^2 \left[ \frac{(1 + x^2)^2}{1 + x_0^2} + \frac{2(1 + x_0^2)^2}{1 + x^2} \right]
\end{equation}
with the initial normalized aspiration length, \(x_0\).

Substituting Eqs.~\ref{eq:bE},~\ref{eq:aE} and ~\ref{eq:v} into Eq.~\ref{eq:total}, the $P\text{-}x$ relationship with two parameters, \(K\) and \(\delta\), is obtained as follows:

\begin{equation}
P = \frac{K}{R_{\mathrm{p}}} \left[ \frac{8x}{1 + x_0^2} + (\delta - 1)\frac{8(1 + x_0^2)x}{(1 + x^2)^3} \right],
\label{eq:fitting}
\end{equation}

\noindent where $\delta$ is a dimensionless parameter;
\begin{equation}
\delta \equiv \frac{4 \pi k}{\pi K R_{\mathrm{p}}^2 (1 + x_0^2)},
\end{equation}

\noindent
showing the initial balance between a whole bending energy and area stretching energy in aspirated regions.
Further, by expanding Eq.~\ref{eq:fitting} to first order in x and considering that \(k\) is proportional to the Young’s modulus \cite{Landau2012}, we deduced a model expression consistent with previous formulations that are solely based on linear elasticity \cite{Theret1988}.

Accordingly, the second derivative of the pressure equation is
\begin{equation}
\frac{d^2 P}{dx^2} = \frac{8K(1 + x_0^2)}{R_{\mathrm{p}}} (1 - \delta) \cdot \frac{6x(3-5x^2)}{(1 + x^2)^5},
\end{equation}
which shows \(\delta\) determines the overall mechanical response under the small deformation, \textit{i.e.}, $x < 0.77$ from \(\frac{6x(3 - 5x^2)}{(1 + x^2)^5} >0\).

Fig.~\ref{fig:model}(b) presents the $R_{\mathrm{p}}P/K-x$ curves for various \(\delta\) values from 0 to 4 using Eq.~\ref{eq:fitting}.
When \(\delta = 1\), the curve becomes linear, but when \(\delta < 1\) (area stretch dominant) and \(\delta > 1\) (bending dominant), the curve becomes convex downward and upward, respectively. This equation demonstrates that it is not the bending energy or the stretching energy alone but rather the ratio between them, denoted as $\delta$, that determines the shape of the $P\text{-}x$ curve. While \(K\) determines the overall height of the curve, $\delta$ characterizes the shape of the curve.

\begin{figure}
\includegraphics[width=1 \linewidth] {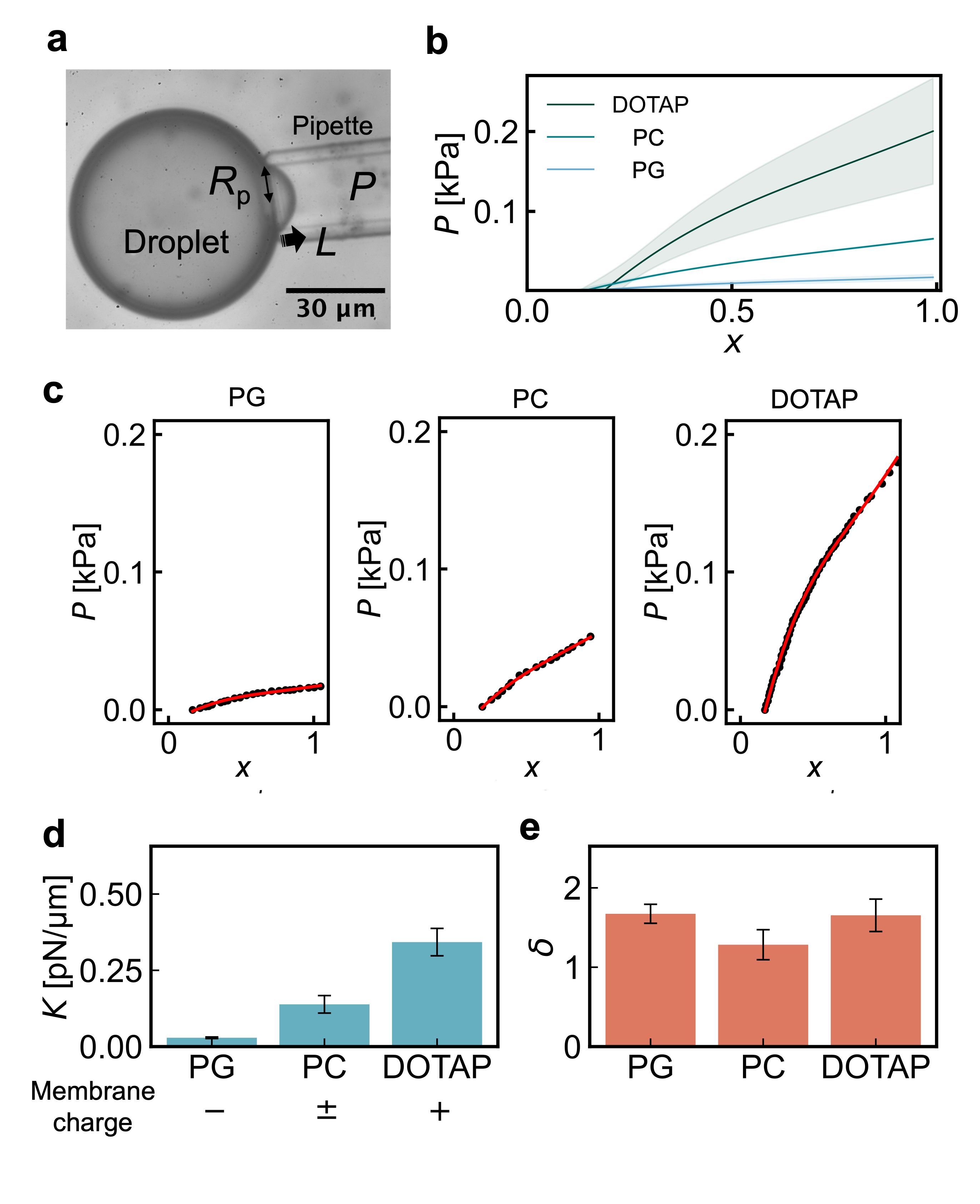}
\caption{\label{fig:microdroplet} Micropipette aspiration of droplets covered with a lipid membrane. (a) A microscope image showing a droplet aspirated by a pipette with $R_{\mathrm{p}}=12$ $\mu$m at an aspiration pressure, \textit{P}. (b, c) The average $P\text{-}x$ curves were obtained by fitting the experimentally obtained curves to Eq. ~\ref{eq:fitting}. The colors indicate the lipid types. (c) An example of the experimentally obtained $P\text{-}x$ curve (shown in dots) and their fitting line (shown in red line). (d, e) The average values of (d) \(K\) and (e) \(\delta\) are derived by the fitting for each lipid. Shaded regions in (b) and the error bars in (d and e) represent the standard deviation ($n$ = 7, 3, and 13 for PG, PC, and DOTAP, respectively).}
\end{figure}

To validate our proposed model, we aspirated the water-in-oil microdroplets covered with a lipid membrane. We selected three different lipids having different elastic properties: 1,2-dioleoyl-sn-glycero-3-phosphocholine (PC, electrically neutral); 1,2-Dioleoyl-3-trimethylammonium-propane (DOTAP, positively charged); 1,2-Dioleoyl-sn-glycero-3-phosphoglycerol (PG, negatively charged). 

 The droplets were prepared using our method \cite{Kurokawa2017}. Briefly, dry lipid films were formed at the bottom of a glass tube and dissolved in hexadecane to a final concentration of 1 mM by sonication for 90 minutes. This lipid-in-oil solution was mixed with 370 mM NaCl in a fixed ratio (volume ratio 1:20), and droplets were prepared by pipetting. To aspirate the droplets with a radius $R> 25$ µm, we used a micropipette with $R_{\mathrm{p}}$ of approximately 15 µm, as shown in Fig.~\ref{fig:microdroplet}(a), following the previously reported method \cite{Sakai2020}. 

To validate Eq.~\ref{eq:fitting}, Fig.~\ref{fig:microdroplet}(b) presents the $P\text{-}x$ curves for droplets coated with three different lipid types: PG, PC, and DOTAP.
Specifically, Fig.~\ref{fig:microdroplet}(c) shows representative examples of individual $P\text{-}x$ curves along with the corresponding fits based on Eq.~\ref{eq:fitting} (solid lines). The fitting curves align with the experimental data with high precision, capturing not only the overall nonlinear trend but also the subtle variations across the full deformation range. The average coefficient of determination, \(R^2\), exceeds 0.99 in all cases, underscoring the model’s exceptional accuracy in fitting the experimental results.

When comparing the $P$ values for these lipids with identical $x$ values, DOTAP shows the highest $P$ value, followed by PC and PG (Fig.~\ref{fig:microdroplet}(c)). Changing the charge of the lipid membrane from negative to positive may enhance the mechanical properties of the droplets.
By fitting the $P\text{-}x$ curves with Eq.~\ref{eq:fitting}, we determined two key mechanical parameters: the area stretching modulus, \(K\), and the energy ratio between bending and area stretching, \(\delta\). As shown in Fig.~\ref{fig:microdroplet}(d), \(K\) is the highest for DOTAP, followed by PC and PG, which aligns with the observed trends in the $P\text{-}x$ curves above. 
In contrast,  \(\delta\) is consistently greater than 1, and no significant differences were found among the lipids (Fig.~\ref{fig:microdroplet}(e)). This result shows that the mechanical properties of the membrane-covered droplets are predominantly governed by stretching rather than bending. It also indicated that the area stretching modulus $K$ increases as the charge of the lipids constituting the membrane transits from negative to positive. In the presence of salt, cationic accumulation in the negatively charged membrane may promote lateral area stretching more than in the positively charged membrane.
To analyze the effect of interior salts, we increased the salt concentration of the NaCl solution from the initial 370 mM to 1110 mM. Interestingly, we observed that the number of DOTAP droplets exhibited convex downward curves ($\delta <1$, see Fig. S1), compared to the upward trends ($\delta >1$) as shown in Fig. 2(b, c). Nevertheless, Eq.~\ref{eq:fitting} remains applicable regardless of the trend direction. This suggests that the contribution of bending to the mechanical response may increase under high-salinity conditions. These results are consistent with previous studies, which indicate that the mechanical properties of lipid vesicles depend on the membrane charge and interior salt conditions \cite{Kato2015}. 

\begin{figure}
\includegraphics[width=1 \linewidth] {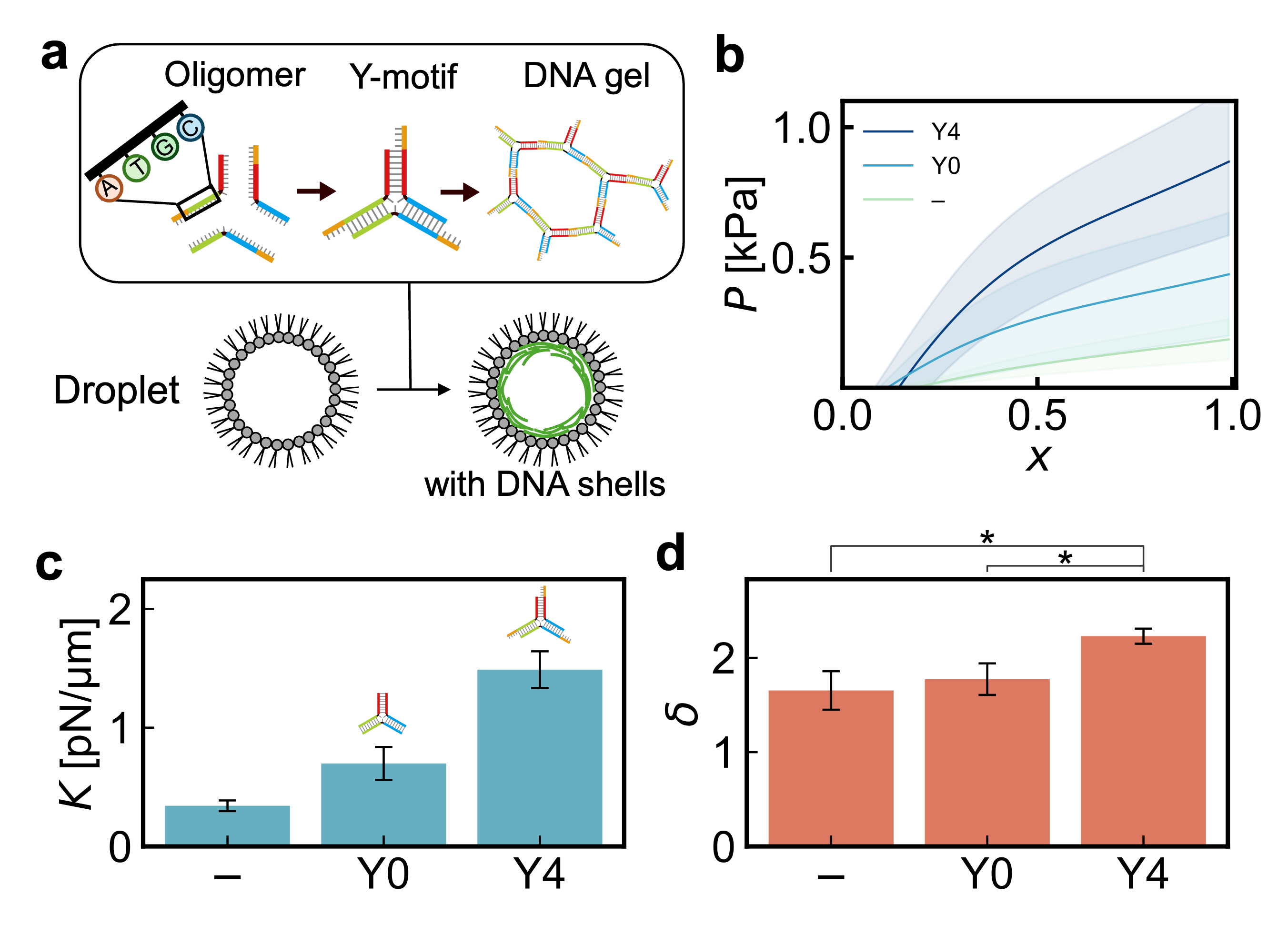}
\caption{\label{fig:microdroplet_DNA} Micropipette aspiration of W/O microdroplets with DNA gel shells. (a) Schematic illustration of DNA hybridization and gelation via annealing of sticky-end strands (orange). The resulting DNA gels localize beneath the membrane due to electrostatic interactions. (b) The average dependence of the normalized aspiration length \(x\) on (\(P\)) for droplets with Y0 and Y4 shells, with the standard deviation indicated in the shaded regions. Values of (c) \(K\) and (d) \(\delta\) for both systems, obtained from the proposed nonlinear elastic model (n=7 for droplets with Y0 shells and n=10 for those with Y4 gel shells). Statistical significance was determined using Welch’s t-test and \(p < 0.05\) (*).}
\end{figure}

We also validated our model using a stiffer shell, which was prepared by DOTAP-coated droplets with a thin shell of DNA gel. As we previously reported, this DNA gel shell, prepared under the DOTAP membrane, mechanically supports the droplets, like cytoskeletal structures of living cells \cite{Kurokawa2017}. Specifically, this DNA gel shell is prepared using three different single-chain DNAs, which spontaneously form Y-motifs and assemble into a gel network by annealing (Fig.~\ref{fig:microdroplet_DNA}(a)). By annealing these DNAs inside DOTAP droplets, we obtained a shell localized underneath the membrane by electrostatic interactions between negatively charged DNA and the cationic lipid DOTAP.

As stiffer gels under the membrane were expected to shift the $P\text{-}x$ curves upward, we used two types of Y motif DNAs, Y4 and Y0. Y4 has a four-base sticky-end capable of gel formation, while Y0 has no sticky-end  and cannot form a gel. Fig.~\ref{fig:microdroplet_DNA}(b) displays how Y4 shifted the $P\text{-}x$ curve upward compared to the bare droplets (control) and the droplets containing Y0. Without surprise, Eq.~\ref{eq:fitting} continues to fit the mechanical response of the stiffer shells (see Fig. S2 for the fitting data of each curve).

The parameters \(K\) and \(\delta\) obtained from the fitting Figs.~\ref{fig:microdroplet_DNA}(c, d) are the highest for the Y4 DNA gel shell. 
These results demonstrate that the DNA gel shell substantially enhances the mechanical robustness of the droplets, both laterally (i.e., increased area stretching with $K$) and perpendicularly (i.e., increased bending with $\delta$). Notably, the extracted values of \(K\) are also comparable in magnitude to those reported in human red blood cells \cite{Lenormand2001}. Furthermore, the higher value of \(\delta\) suggests that gel formation increases the bending contribution in the total elastic energy. Compared to the control, \(K\) is higher for Y0, suggesting that the membrane coating with DNA nanostructures increases the lateral resistance. 
\begin{figure}
\includegraphics[width=1 \linewidth] {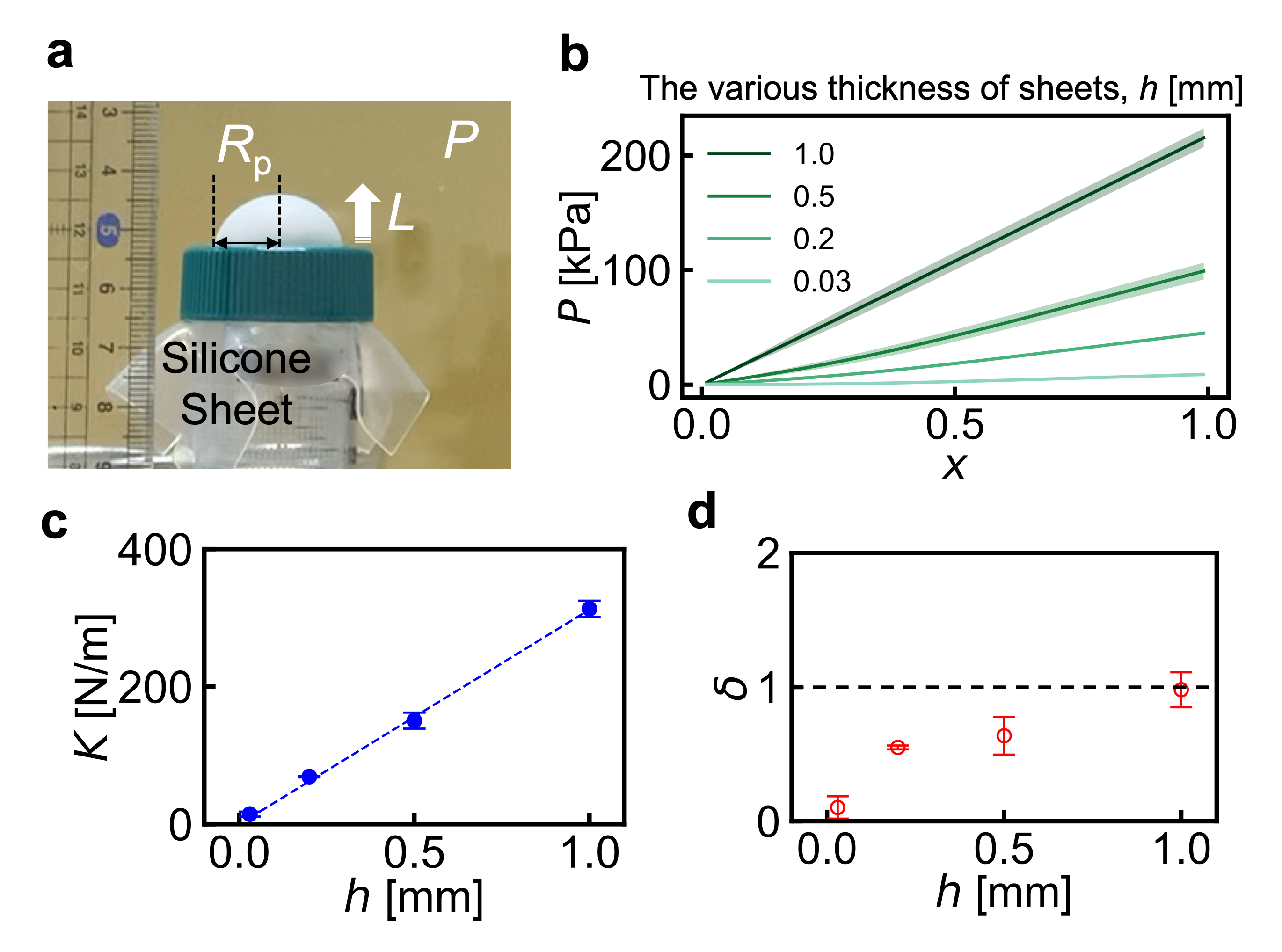}
\caption{\label{fig:silicone} Macroscopic deformation of a silicone sheet fixed to a tube in a desiccator. (a) Experimental setup: A silicone sheet of thickness \( h \) (mm) is secured to a plastic tube (\( R_{\mathrm{p}} = 11.5 \) mm) and subjected to a pressure difference (\( P \)). (b) Averaged \( P\text{-}x \) curves for sheets of various $h$. The light-colored areas represent their SD values (n=3 for \(h=0.03\), n=4 for \(h=0.\), n=3 for \(h=0.5\) and n=3 for \(h=1.0\)). (c, d) The $h$ dependence of the derived average values of (c) \(K\) and (d) \(\delta\). The dashed line in (d) shows \(\delta=1\), where the curve is linear.} 
\end{figure}
To show the versatility of Eq.~\ref{eq:fitting} across different length scale deformations, we experimentally analyzed the $P\text{-}x$ curves for silicone sheets fixed on a tube inside a desiccator, as shown in Fig.~\ref{fig:silicone}(a). 
It is to be noted that the number of molecules in the inflated region remains nearly constant throughout the experiment. The boundary of the deforming region is also fixed in our setup to match closely to our aspiration experiments.
We commercially obtained a silicone sheet with a thickness \(h\) of 0.03, 0.2, 0.5, and 1.0 mm. Analogous to the micropipette aspiration setup (Fig.~\ref{fig:microdroplet}(a)), a pressure difference, $P$, was induced using a desiccator, and the resulting aspiration length $L$ was captured. Despite transitioning from microscopic to macroscopic scales, Eq.~\ref{eq:fitting} fits the overall $P\text{-}x$ curves for all sheets with various $h$ (see Fig. S3). 

Fig.~\ref{fig:silicone}(b) shows that for all sheets of different thicknesses, the $P\text{-}x$ curves are consistently downward convex, in contrast to the upward convex $P\text{-}x$ curves for membrane-covered droplets (Figs. ~\ref{fig:microdroplet} and ~\ref{fig:microdroplet_DNA}). 

As shown in Fig.~\ref{fig:silicone}(c), the value of \(K\) obtained by fitting increases linearly with \(h\), consistent with classical elastic theory, which states that \(K \propto h\) \cite{Landau2012, Li2011}. Additionally, the value of \(\delta\) increases with \(h\) and approaches unity (Fig.~\ref{fig:silicone}(d)), indicating ideal elastic behavior. These results demonstrate that our proposed model using \(K\) and \(\delta\), effectively evaluates the $P\text{-}x$ curves and assesses mechanical properties in the macroscale. 

In summary, we introduced a physics-based analytical model that captures the nonlinear mechanical response observed during aspiration-induced deformation of elastic shells. This model couples bending and in-plane stretching deformations through two distinct parameters, \(K\) and \(\delta\), which are crucial for understanding the mechanics of various soft matter systems. By fitting the overall \(P\text{–}x\) curves, the model provides enhanced accuracy compared to traditional linear analyses, which often assume homogeneous, linearly elastic behavior. Our experimental results validate the versatility of the proposed model across micro-to-macro scales. Together, these results establish a new framework for probing and interpreting the mechanical properties of elastic sheets and shells, overcoming the limitations of conventional linear models in various soft matter systems, including polymers, hydrogels, and biologically relevant materials.

$ Acknowledgments- $ This research was partially funded by the Japan Society for the Promotion of Science (JSPS) KAKENHI (grant numbers  22H01188, 24H02287 (M.Y.), the Japan Science and Technology Agency (JST) (grant numbers FOREST, JPMJFR213Y; CREST (JPMJCR22E1) (M. Y.)), and the
World-Leading Innovative Graduate Study Program for Advanced Basic Science Course (WINGS-ABC) at the University of Tokyo (K. M.). We thank Mr. Daisuke S. Shimamoto and Dr. Anusuya Pal for valuable comments. 
\nocite{*}

\bibliography{main}% Produces the bibliography via BibTeX.

\end{document}